\documentclass[11pt,twoside,letterpaper,reqno]{amsart}
\usepackage{amsfonts,amsmath,amsthm,amssymb}
\usepackage[pdftex]{graphicx}
\usepackage{hyperref}%\urlstyle{same}
\usepackage[margin=2.5cm]{geometry}

\theoremstyle{plain}
\newtheorem{theorem}{Theorem}%[section]
\newtheorem{lemma}[theorem]{Lemma}

\theoremstyle{definition}

\def\DEFINE#1{\emph{#1}}                     % definitions

\makeatletter
\def\begin@lgo{
    \begin{minipage}{1in}\begin{tabbing}
          \qquad\=\qquad\=\qquad\=\qquad\=\qquad\=\qquad\=\qquad\=\kill
}
\def\end@lgo{\end{tabbing}\end{minipage}}

\makeatother

\def\scan{\textsc{Scan}}

\def\even{\mathit{Even}}
\def\odd{\mathit{Odd}}
\def\etal{\textsl{et al.\@}}
\newcommand{\cells}{\ensuremath{\mathcal{C}}}

\def\paren#1{\ensuremath{\left({#1}\right)}}
\def\bracket#1{\ensuremath{\left[{#1}\right]}}

\def\ceil#1{\ensuremath{\left\lceil{#1}\right\rceil}}

\def\br#1{\ensuremath{\beta(#1)}}

\newcommand{\leftdangle}{\ensuremath{\langle\!\langle}}
\newcommand{\rightdangle}{\ensuremath{\rangle\!\rangle}}
\def\note#1#2{\begin{quote}\marginpar{$\bigstar$}\textsf{\leftdangle{#1:~#2}\rightdangle}\end{quote}}

\newcommand{\myomit}[1]{}
\renewcommand{\le}{\leqslant}
\renewcommand{\ge}{\geqslant}
\renewcommand{\leq}{\leqslant}
\renewcommand{\geq}{\geqslant}
\newcommand{\Xmin}{X_{1}}
\newcommand{\Xmax}{X_{2}}
\newcommand{\Ymin}{Y_{1}}
\newcommand{\Ymax}{Y_{2}}

\begin{document}
%---
\title{Cache-Oblivious Selection in Sorted $X+Y$ Matrices}

\author{Mark de Berg}
\address{Mark de Berg\\
         Department of Computer Science\\
         Technische Universiteit Eindhoven\\
         The Netherlands}
\email{mdberg@win.tue.nl}

\author{Shripad Thite}
\address{Shripad Thite\\
         California Institute of Technology\\
         Center for the Mathematics of Information\\
         USA}
\email{shripad@caltech.edu}

%\date{Last revised January 28, 2008}

%% ##################################################################
\begin{abstract}
  Let $X[0 .. {n-1}]$ and $Y[0 .. {m-1}]$ be two sorted arrays, and
  define the $m \times n$ matrix~$A$ by $A[j][i] = X[i] + Y[j]$.
  Frederickson and Johnson~\cite{frederickson84generalized} gave an
  efficient algorithm for selecting the $k$th smallest element
  from~$A$.  We show how to make this algorithm IO-efficient.  Our
  cache-oblivious algorithm performs $O((m+n)/B)$ IOs, where $B$ is
  the block size of memory transfers.
\end{abstract}
%% ##################################################################

\maketitle

%%%%%%%%%%%%%%%%%%%%%%%%%%%%%%%%%%%%%%%%%%%%%%%%%%%%%%%%%%%%%%%%%%%%%%%%%%%%%
%%%%%%%%%%%%%%%%%%%%%%%%%%%%%%%%%%%%%%%%%%%%%%%%%%%%%%%%%%%%%%%%%%%%%%%%%%%%%
%%%%%%%%%%%%%%%%%%%%%%%%%%%%%%%%%%%%%%%%%%%%%%%%%%%%%%%%%%%%%%%%%%%%%%%%%%%%%
%%%%%%%%%%%%%%%%%%%%%%%%%%%%%%%%%%%%%%%%%%%%%%%%%%%%%%%%%%%%%%%%%%%%%%%%%%%%%
\section{Introduction}

Let $S$ be a multi-set of elements from a totally ordered universe and
let $k$ be an integer in the range $1 \le k \le |S|$.  The
\emph{selection problem} is to find a $k$th smallest element of $S$,
that is, an element $x \in S$ that is $k$th in some non-decreasing total
ordering of~$S$. Selection is a fundamental problem in computer
science and a key building block of many algorithms.  Selection
is trivial when $S$ is sorted, but when $S$ is not given in sorted
order it becomes more challenging.  A classical divide-and-conquer
algorithm~\cite{blum73selection,cormen01algorithms} solves the
selection problem for unsorted inputs in $O(|S|)$ time.

Often, the input is naturally organized as a two-dimensional matrix~$A$
with $m$ rows and $n$ columns. Using the classical algorithm one can
perform selection in $A$ in $O(mn)$ time, which is optimal in
the worst case.  When the rows and columns of the matrix are sorted,
however, one can do much better. Frederickson and
Johnson~\cite{frederickson84generalized,frederickson90erratum} gave an
algorithm for this case---we will call it the \emph{FJ-algorithm} from now on---that
runs in  $O(m\lg (2n/m))$ time; here we assume without loss of generality
that $m \le n$. Note that when $m=n$ the running time is
simply $O(n)$.

In some applications the matrix~$A$ is defined succinctly by the
Cartesian product of two given vectors $X[0 .. {n-1}]$ and $Y[0
.. {m-1}]$. We are interested in the case where $A=X+Y$, that is,
\[
  A[j][i] = X[i] + Y[j]
\]
where $X$ and $Y$ are sorted.  (The symbol `+' can mean any monotone
binary operator.)  Since $X$ and $Y$ are sorted, the rows and columns
of $A$ are sorted.  Hence, one can perform selection in $A$ in $O(m
\lg (2n/m))$ time by FJ-algorithm.
Selection in such sorted $X+Y$ matrices is used as a subroutine
in several other algorithms---see~\cite{agarwal98efficient,bdkst-cmotc-98,ek-cfbm-96,gks-gsops-95,sk-lrm-99,sw-rpp-96} for some examples.

The FJ-algorithm is efficient in terms of CPU computation time.
Unfortunately, it is not efficient when it comes to IO
behavior, because it accesses elements of the input
arrays $X$ and $Y$ non-sequentially, in a pattern that does not
exhibit locality of reference.
This is the goal of our paper: to develop
a variant of the algorithm that has better IO behavior.
\medskip

The \emph{input-output complexity}, or \emph{IO-complexity}, of an
algorithm is usually analyzed in the external-memory model introduced
by Aggarwal and Vitter~\cite{aggarwal88iocomplexity}.  In this model
the memory consists of two levels: a fast memory and a slow
memory. The fast memory can store up to $M$ words and the slow memory
has unlimited storage capacity.  Data is stored in the slow memory in
blocks of size~$B$. To be able to do computations on data in the slow memory,
that data first has to be brought into the fast memory;
data which is evicted from fast memory (to make room for other data)
needs to be written back to the slow memory. Data is transferred
between fast and slow memory in blocks.  The IO-complexity of an
algorithm is the number of block transfers it performs.

The two levels in this abstract model can stand for any two
consecutive levels in a multi-level memory hierarchy: the slow memory
could be the disk and the fast memory the main memory, the slow memory
could be the main memory and the fast memory the L3 cache, and so on.
The values of $M$ and $B$ are different at
different levels; the higher up in the memory hierarchy, the larger
the memory size $M$ and block size $B$.
\medskip

Our main result is a variant of the FJ-algorithm for sorted $X+Y$ matrices whose
IO-complexity is $O(\scan(n+m))$.  Here, $\scan(s)$ is the number of
IOs performed when scanning~$s$ consecutive items; $\scan(s) \le 1 +\ceil{s/B}$.
Our algorithm is \emph{cache-oblivious}~\cite{frigo99cacheoblivious}, which means it
is oblivious of the parameters~$M$ and~$B$.
%Thus, the algorithm cannot
%explicitly do its own memory management.  For example, it cannot
%decide which items to put together into one block because it does not
%know the block size~$B$.
In other words, the parameters $M$ and $B$ are only used in the analysis
of the algorithm; they are not used in the algorithm itself.
The beauty of cache-oblivious algorithms is that, since they do not
depend on the values $M$ and $B$, they are IO-efficient for all values
of~$M$ and~$B$ and, hence, IO-efficient at all levels of a multi-level
memory hierarchy.\footnote{In the analysis of cache-oblivious algorithms it is assumed that the
operating system uses an optimal block replacement strategy---see the
paper by Frigo~\etal{}~\cite{frigo99cacheoblivious} for a
justification of this and some other assumptions in the model.}

%%%%%%%%%%%%%%%%%%%%%%%%%%%%%%%%%%%%%%%%%%%%%%%%%%%%%%%%%%%%%%%%%%%%%%%%%%%%%
%%%%%%%%%%%%%%%%%%%%%%%%%%%%%%%%%%%%%%%%%%%%%%%%%%%%%%%%%%%%%%%%%%%%%%%%%%%%%
%%%%%%%%%%%%%%%%%%%%%%%%%%%%%%%%%%%%%%%%%%%%%%%%%%%%%%%%%%%%%%%%%%%%%%%%%%%%%
%%%%%%%%%%%%%%%%%%%%%%%%%%%%%%%%%%%%%%%%%%%%%%%%%%%%%%%%%%%%%%%%%%%%%%%%%%%%%
\section{The FJ-Algorithm}

First, we give a rough outline of the
FJ-algorithm~\cite{frederickson84generalized}. A detailed description
is given in Figure~\ref{fig:algFJ}.

Let $X[0 .. {n-1}]$ and $Y[0 .. {m-1}]$ be two input arrays of real numbers,
given in sorted order:~$X[0] \le X[1] \le \cdots \le X[n-1]$ and $Y[0]
\le Y[1] \le \cdots \le Y[m-1]$.  Let $A[0 .. {m-1}][0 .. {n-1}]$ be the
matrix $X+Y$, that is, the matrix defined by $A[j][i]=X[i]+Y[j]$.  We
assume that $m = n$ and that $n$ is a power of~2; this can easily be
ensured by implicitly padding the arrays $X$ and $Y$ suitably.

Following Frederickson and Johnson, we call a submatrix of $A$ a
\emph{cell}.  The algorithm maintains a set $\cells$ of \emph{active
cells}, such that the desired element will be present in one of the
active cells.  Initially, the entire matrix $A$ is the sole active
cell.

The algorithm proceeds in $\lg n$ iterations.  Let $\cells_p$ denote
the set of active cells at the beginning of the $p$th iteration, where
$p=1$,$2$,$\ldots$,$\lg n$. The $p$th iteration begins by splitting
each cell of $\cells_p$ into four smaller cells by bisecting each
dimension. Let $\cells^*_p$ denote the list of cells obtained by
splitting each cell of $\cells_p$ into four.  The algorithm next
discards certain cells from $\cells^*_p$ which do not contain the
desired element, thus obtaining the set $\cells_{p+1}$ to be used in
the next iteration.

Cells are discarded based on their minimum and maximum elements.  A
cell $C\in \cells^*_p$ for which $\min(C)$ is larger than a certain
number of other minima can safely be discarded because all elements of
$C$ will be larger than the desired element.  Similarly, a cell $C\in
\cells^*_p$ for which $\max(C)$ is smaller than a certain number of
other maxima can be discarded because all elements of $C$ will be
smaller than the desired element.  The exact condition for discarding
cells is given in step~(\ref{stepb}) of the algorithm in
Figure~\ref{fig:algFJ}.

The cells in $\cells_p$ have size
$(n/2^{p-1})\times(n/2^{p-1})$ and the cells in
$\cells^*_p$ have size $(n/2^{p})\times(n/2^{p})$.  Hence,
after iteration $p=\lg n$, the cells in $\cells_p$ are singletons
(that is, $1\times 1$ cells).  The classical selection algorithm is
then used to find the desired element among these singletons.

%% ................................................................
\begin{figure*}[t]
\hrule
\begin{center}
\textsc{FJ-algorithm}$(X,Y,k)$:
\begin{enumerate}
  \item \label{step1} Initialize $\cells_1$ such that its only cell is
the entire matrix $A = X+Y$.
  \item \label{step2} {\bf for} $p := 1$ {\bf to} $\lg n$ {\bf do}
        \begin{enumerate}
        \item \label{stepa} Split each $C\in\cells_p$ into four subcells to
              obtain the set~$\cells^*_p$.
              Let $L_p := \min\{n,2^{p+1}-1\}$.
        \item \label{stepb} Let $q := \ceil{k 4^p/n^2} + L_p$. \\
              {\bf if} $q \le |\cells_p^*|$ \\
              {\bf then} \begin{minipage}[t]{125mm}
                         Use a standard selection algorithm to select a $q$th element $x_u$
                         in the multiset $\{\min(C) : C\in \cells^*_p\}$.
                         Discard $|\cells^*_p|-q+1$ cells from $\cells^*_p$, retaining every cell $C$
                         with $\min(C) < x_u$ and no cell with $\min(C)>x_u$.
                         \end{minipage}
        \item \label{stepc} Let $r := \ceil{k 4^p/n^2} - L_p$. \\
              {\bf if} $r\ge 1$ \\
              {\bf then} \begin{minipage}[t]{125mm}
                         Use a standard selection algorithm to select
                         an $r$th element $x_l$ in the multiset $\{\max(C) : C\in \cells^*_p\}$.
                         Discard $r$ cells from $\cells^*_p$, retaining every cell $C$
                         with $\max(C) < x_l$ and no cell with $\max(C)>x_l$.
                         \end{minipage}
        \item \label{stepd} Let $k := k - r(n^2/4^p)$ and let $\cells_{p+1} := \cells^*_p$.
        \end{enumerate}
  \item \label{step3} Select the $k$th element from the cells in $\cells_p$ using a standard selection algorithm.
\end{enumerate}
\end{center}
\hrule
\caption{The matrix selection algorithm of Frederickson and Johnson~\cite{frederickson84generalized}.}
\label{fig:algFJ}
\end{figure*}
%% ................................................................

The following theorem stating the performance of the
FJ-algorithm is a special case of the general theorem proved by
Frederickson and Johnson~\cite{frederickson84generalized}.

\begin{theorem}{\cite{frederickson84generalized}}
  Given two sorted arrays $X$ and $Y$, each of size~$n$,
  the FJ-algorithm correctly computes an element of rank $k$ in the
  matrix $A = X + Y$ in $O(n)$ time.
\end{theorem}

%%%%%%%%%%%%%%%%%%%%%%%%%%%%%%%%%%%%%%%%%%%%%%%%%%%%%%%%%%%%%%%%%%%%%%%%%%%%%
%%%%%%%%%%%%%%%%%%%%%%%%%%%%%%%%%%%%%%%%%%%%%%%%%%%%%%%%%%%%%%%%%%%%%%%%%%%%%
%%%%%%%%%%%%%%%%%%%%%%%%%%%%%%%%%%%%%%%%%%%%%%%%%%%%%%%%%%%%%%%%%%%%%%%%%%%%%
%%%%%%%%%%%%%%%%%%%%%%%%%%%%%%%%%%%%%%%%%%%%%%%%%%%%%%%%%%%%%%%%%%%%%%%%%%%%%
\section{IO-Efficient Selection}

Next, we show how to make the algorithm of the previous section
IO-efficient.  Henceforth, we will refer to the slow memory in our
two-level hierarchy as the \emph{disk} and to the fast memory as the
\emph{cache}.  We assume that the array~$X$ is laid out in order
in~$n$ consecutive memory locations on disk.  Similarly, the array~$Y$
is laid out in order in~$n$ consecutive memory locations on disk.

The FJ-algorithm needs an efficient selection algorithm
in steps~(\ref{stepb}),~(\ref{stepc}), and~(\ref{step3}). Fortunately,
the standard selection algorithm has good IO-behavior.

%% ..............................................................
\begin{lemma}
\label{le:standard-selection}
  The standard selection algorithm~\cite{blum73selection}
  selects an element of a given rank~$k$ from an array of~$s$
  elements in $O(s)$ time and using $O(\scan(s))$ IOs.
\label{lemma:standard-selection}
\end{lemma}
%% ..............................................................

%We could apply Lemma~\ref{lemma:standard-selection} to select the
%desired element in the matrix $A$ in $O(\scan^2(n))$ IOs and $O(n^2)$
%time. However, this would be vastly inferior to the
%%and Johnson which we develop next.

Even though selection is the main subroutine used by the matrix
selection algorithm, Lemma~\ref{lemma:standard-selection} does not
imply that the FJ-algorithm is IO-efficient.  The main
problem is that maintaining the list of active cells can dominate the
IO-cost of a na\"ive implementation of the FJ-algorithm,
leading to $O(n)$ IO-complexity rather than $O(\scan(n))$.
To make the algorithm IO-efficient, we need
to take a detailed look at the manipulation of active cells.

The FJ-algorithm needs a data structure to store the
sets $\cells_p$ and $\cells_p^*$ of active cells. One could use
linked lists, but traversing a linked list is not
IO-efficient because adjacent list elements could be stored in
different blocks, requiring as many as one IO-operation per list
element.  Instead, we use arrays, which can store any list $L$
compactly on disk in $O(|L|/B)$ blocks.

We represent a cell $A[j_1..{j_2-1}][i_1..{i_2-1}]$ by the $8$-tuple
\[
  \paren{ i_1, j_1, i_2, j_2, X[i_1], X[i_2-1], Y[j_1], Y[j_2-1] },
\]
and we identify a cell with its corresponding $8$-tuple.
The active cells are stored in lexicographic order of their corresponding
$8$-tuples.
From the $8$-tuple representing cell $C$ we can compute
$\min(C)=X[i_1]+Y[j_1]$ and $\max(C)=X[i_2-1]+Y[j_2-1]$ in $O(1)$ time
and no additional IOs.  Hence, steps~(\ref{stepb}),~(\ref{stepc}),
and~(\ref{step3}) of the FJ-algorithm can all be done in
$O(\scan(|\cells^*_p|))$ IOs. The problem lies in step~(\ref{stepa}),
where we compute $\cells_p^*$ from $\cells_p$ by splitting each cell
into four subcells.
%Maintaining the lexicographic order, when computing $\cells^*_p$ by splitting
%cells in $\cells_p$, is not trivial, because subcells of two disjoint
%cells are interleaved in the lexicographic order.

Suppose we have to split the cell $\paren{i_1, j_1, i_2, j_2, X[i_1],
X[i_2], Y[j_1], Y[j_2]}$.
Let $i_m = (i_1 + i_2)/ 2$ and let $j_m = (j_1 + j_2)/2$.
The four subcells we must generate are as follows:  % See Figure~\ref{fig:split}.
\\[2mm]
\includegraphics{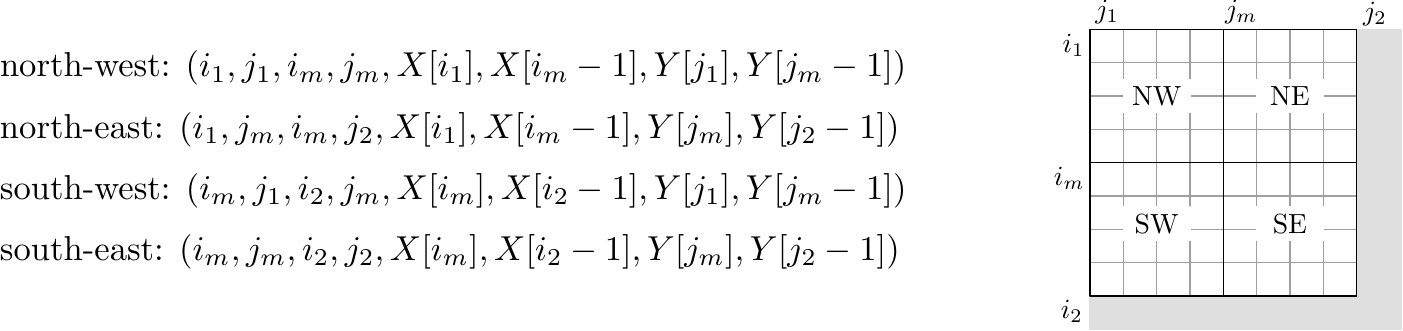}
\\
%\begin{enumerate}
%\item $\paren{i_1, j_1, i_m, j_m, X[i_1], X[i_m-1], Y[j_1], Y[{j_m}-1]}$ \\[-2mm]
%\item $\paren{i_1, j_m, i_m, j_2, X[i_1], X[i_m-1], Y[j_m], Y[j_2-1]}$ \\[-2mm]
%\item $\paren{i_m, j_1, i_2, j_m, X[i_m], X[i_2-1], Y[j_1], Y[j_m-1]}$ \\[-2mm]
%\item $\paren{i_m, j_m, i_2, j_2, X[i_m], X[i_2-1], Y[j_m], Y[j_2-1]}$ \\[-2mm]
%\end{enumerate}
%
Most components of the subcells can be computed from the components of
$C$, except that $X[i_m-1]$ and $X[i_m]$ need to be fetched from the
array~$X$, and $Y[j_m-1]$ and $Y[j_m]$ need to be fetched from the
array~$Y$.  If we are not careful, fetching these values will cost us
an IO each time and the whole algorithm will not be IO-efficient. Next
we describe how to overcome this problem.
\medskip

Let us examine in what order the algorithm accesses the array $X$; the
array $Y$ will be discussed later.

First consider the elements from $X$ needed for the fifth component of the
cells, which stores the minimum $X$-value in the cell.
In the initialization step, the entire matrix $A$ is the only active
cell; its minimum $X$-value is  $X[0]$. In the first iteration ($p=1$) we split $A$
into four subcells. The minimum $X$-values in those subcells
are either $X[0]$ (for the north-west and south-west subcells) or $X[n/2]$
(for the north-east and south-east subcells). Since $X[0]$ is conveniently
stored in the original cell, we only need to access $X[n/2]$.
In the second iteration, we need to access $X[n/4]$ and $X[3n/4]$.
In general, in the $p$th iteration ($1\le p<\lg n$), each active cell in $\cells_p$ has dimension
$(n/2^{p-1}) \times (n/2^{p-1})$, and the elements that need to be
accessed to obtain their minimum X-values are
$X[(2i-1) \cdot (n/2^p)]$ for $1\le i\le 2^{p-1}$.
(In fact, we do not necessarily need all these elements,
since not all cells have to be active.)

To enable IO-efficient access to these elements in $X$, we construct an array $\Xmin[1..n/2-1]$
that stores, for any $p$ with $1\leq p<\lg n$,
the elements needed in the $p$th iteration consecutively.
Thus we define array $\Xmin$ so that it has the
following property:
%% .................................................................
\begin{quotation}
For all $p$ in the range $1 \le p < \lg n$,
for all $i$ in the range $1 \le i \le 2^{p-1}$,
we have
\begin{equation}
\Xmin\bracket{2^{p-1} + i-1} = X\bracket{(2i-1) \frac{n}{2^p}}.  \label{def:permutation1}
\end{equation}
\end{quotation}
%% .................................................................
Note that, together with $X[0]$, the elements in $\Xmin$ are
exactly the elements in $X$ at even-numbered positions.

Similarly, the elements that need to be
accessed to obtain the maximum X-values in the $p$-th iteration, namely
$X[(2i-1) \cdot (n/2^p-1)]$ for $1\le i\le 2^{p-1}$,
are stored in an array $\Xmax$.
Thus array $\Xmax[1..n/2-1]$ stores the odd-numbered elements in $X$
(except $X[n-1]$), as follows:
%% .................................................................
\begin{quotation}
  For all $p$ in the range $1 \le p < \lg n$, for all $i$ in the
  range $1 \le i \le 2^{p-1}$, we have
\begin{equation}
  \Xmax\bracket{2^{p-1} + i-1} = X\bracket{(2i-1)\,\frac{n}{2^p} - 1}. \label{def:permutation2}
\end{equation}
\end{quotation}
%% .................................................................
Next we show how to compute the array $\Xmin$ efficiently;
$\Xmax$ can be computed similarly.
\medskip

Given an integer $i$, the \emph{bit-reversal}
of $i$ is the integer $\br{i}$ such that the binary string
representing $\br{i}$ is the reverse of the binary string
representing~$i$.  The \emph{bit-reversal permutation} $Z'$ of an array $Z$
is the permutation that maps that $Z[i]$ to $Z'[\br{i}]$. The
bit-reversal permutation can be computed recursively as follows: Copy
all elements in even-numbered positions in $Z$ in order to the first
half of the array $Z'$, and copy all elements in odd-numbered positions
of $Z$ in order to the second half of $Z'$; recurse on both halves.

Now suppose we only recurse on the first half of
the array $Z'$; the elements in the second half
are kept in the same relative order as in the input array~$Z$.
We call the resulting permutation the \emph{partial bit reversal}.
As we will show below, the partial bit reversal of array $X$
is closely related to the array $\Xmin$ that we want to compute.
The recursive algorithm PBR given in Fig.~\ref{fig:algPBR}---a non-recursive
version would also be possible---computes
a partial bit reversal $Z'$ of a given array $Z[0..n-1]$.
In the initial call, $Z'$ is a copy of $Z$, and $s=n$.
(Recall that we assumed $n$ is a power of~2.)

\begin{figure}[h]
\hrule
\begin{center}
\textsc{Algorithm PBR}$(Z',s)$:
\begin{enumerate}
  \item {\bf if} $s>1$
  \item \label{pbr2} {\bf then} \begin{minipage}[t]{125mm}
                     \textsl{Comment: $\even[0..s/2-1]$ and $\odd[0..s/2-1]$ are auxiliary arrays.}
                     \begin{enumerate}
                     \item \label{pbr2a} {\bf for} $i := 0$ {\bf to} $s-1$ {\bf do} \\
                           \hspace*{3mm} {\bf if} $i$ is even
                                         {\bf then} $\even[i/2] := Z'[i]$
                                         {\bf else} $\odd[(i-1)/2]:= Z'[i]$
                     \item \label{pbr2b} {\bf for} $i := 0$ {\bf to} $s/2-1$ {\bf do} $Z'[i] := \even[i]$ \\
                                         {\bf for} $i := s/2$ {\bf to} $s-1$ {\bf do} $Z'[i] := \odd[i-s/2]$
                     \item \label{pbr2c} PBR$(Z',s/2)$
                     \end{enumerate}
                   \end{minipage}
\end{enumerate}
\end{center}
\hrule
\caption{Algorithm to compute a partial bit-reversal permutation}
\label{fig:algPBR}
\end{figure}
The following lemma, which gives the running time and IO complexity of PBR,
follows easily from the fact that
steps~\ref{pbr2a}~and~\ref{pbr2b} of PBR are just linear scans of arrays
$Z'$, $\even$, and $\odd$, so these steps run in $O(s)$ time and $O(\scan(s))$ IOs.

%% ..................................................................
\begin{lemma}
  PBR$(Z',n)$ runs in $O(n)$ time and uses $O(\scan(n))$ IOs.
\label{lemma:permutecost}
\end{lemma}
The next lemma shows the correspondence between the partial bit reversal
of our input array $X$ and the array $X_1$ we want to compute.
It implies that  $\Xmin$ can be obtained by computing the partial bit
reversal $X'$ of $X$ and then taking the elements from $X'[1..n/2-1]$ in order.

%% ...................................................................
\begin{lemma}\label{le:reorder}
Let $Z'$ be the partial bit reversal of an array $Z[0..n-1]$,
where $n$ is a power of 2, as computed by PBR.
Then for all $p$ in the range $1 \le p < \lg n$,
for all $i$ in the range $1 \le i \le 2^{p-1}$,
we have
\[
Z'\bracket{2^{p-1} + i-1} = Z\bracket{(2i-1) \frac{n}{2^p}}.
\]
\label{lemma:permutecorrectly}
\end{lemma}
\begin{proof}
Let $j>0$ be an even index, and let $\ell\geq 1$ and $k\geq 1$ be such that
$j=(2\ell-1)2^k$. Now consider what happens to element $Z[j]$.
Initially $Z'$ is a copy of $Z$, so $Z[j]$ is stored in $Z'[j]$. Then,
in the first call to PBR---that is, the call with $s=n$---it
will be moved to $Z'[j/2]$ by steps~(\ref{pbr2a})~and~\ref{pbr2b}.
In the recursive call with $s=n/2$ it will be moved to $Z'[j/4]$
(if $k>1$).
This process continues $k$ times, until the recursive call is
made with $s=n/2^k$. At this point $Z[j]$ is stored in $Z'[2\ell-1]$,
and step~(\ref{pbr2a}) moves the element to $Z'[n/2^{k+1}+\ell-1]$.
After that it will not be moved anymore by the algorithm.

Now set $i=\ell$ and take $p$ such that $2^k=n/2^p$. Then
$n/2^{k+1}=2^{p-1}$ and we can conclude that
$Z[(2i-1)\cdot(n/2^p)]$ ends up in $Z'[2^{p-1} + i-1]$, as required.
\end{proof}
%
%\mycomment{The appendix gives an alternative proof using an invariant. But I
%think it is needlessly complicated, less intuitive, and in fact less easy to verify the correctness.}
%
In what follows, we use $\beta_1(j)$ to denote the position of
$X[j]$ in the array $X_1$, for $j>0$ and $j$ even. Thus, according to Equation~(\ref{def:permutation1}), we
have $\beta_1((2i-1) \cdot(n/2^p)) = 2^{p-1} + i-1$. Similarly,
$\beta_2(j)$ denotes the position of
$X[j]$ in the array $X_2$, for $j<n-1$ and $j$ odd; thus
$\beta_2((2i-1) \cdot(n/2^p)-1) = 2^{p-1} + i-1$.
\medskip

The arrays $\Xmin$ and $\Xmax$ give us the $X[\cdot]$-values in the order
they are needed by the cell-partitioning step of the
FJ-algorithm.  However, to partition a cell we also
need to fetch new $Y[\cdot]$ values.  For this we would like to use
the same approach: compute in a preprocessing step two arrays
$\Ymin[1..n/2-1]$ and $\Ymax[1..n/2-1]$, which contain the
$Y[\cdot]$-values in the order needed by the algorithm.  With the
$X[\cdot]$-values this approach was possible, because the cells in
$\cells_p$ are kept in lexicographical order, with the $i_1$-value
being dominant. Hence, we knew exactly not only which
$X[\cdot]$-values were needed in the $p$-th iteration (namely
$X[(2i-1)\cdot (n/2^{p})]$ for $1\le i\le 2^{p-1}$), but also in
which order (namely according to increasing index).  But for the
$Y[\cdot]$-values we only know which values we need in the $p$-th
iteration; we do not know in which order we need them, because the $i_1$-coordinate is
dominant in the order of the cells in $\cells_p$.  Next we will show
that the approach works nevertheless.  Thus we compute arrays $\Ymin$
and $\Ymax$ in exactly the same way as the arrays $\Xmin$ and $\Xmax$ were
computed. Then we partition the cells with the algorithm shown in
Figure~\ref{fig:partition}.

%% ...................................................................
\begin{figure*}[t]
\hrule
\begin{center}
\textsc{Partition}$(\cells_p,\Xmin,\Xmax,\Ymin,\Ymax,p)$:
\begin{enumerate}
  \item Let $\cells_{p,R}$ and $\cells_{p,L}$ be two arrays of twice the size as $\cells_p$.
  \item \label{halveonx-2} {\bf for} $i := 0$ {\bf to} $|\cells_P|-1$ {\bf do} \\
        \begin{minipage}[t]{135mm}
             Let $C=\paren{i_1,j_1,i_2,j_2,X[i_1],X[i_2-1],Y[j_1],Y[j_2-1]}$
               be the cell in $\cells_p[i]$. \\
             Let $i_m = (i_1 + i_2)/ 2$ and let $j_m = (j_1 + j_2)/2$.
             \begin{enumerate}
             \item Fetch $X[i_m-1]$ from $\Xmin[\beta_1(i_m-1)]$ and $X[i_m]$ from $\Xmax[\beta_2(i_m)]$ \label{partition.fetchx}
             \item Fetch $Y[j_m-1]$ from $\Ymin[\beta_1(j_m-1)]$ and $Y[j_m]$ from $\Ymax[\beta_2(j_m)]$ \label{partition.fetchy}
             \item $\cells_{p,L}[2i] \leftarrow \paren{i_1,j_1,i_m,j_m,X[i_1],X[i_m-1],Y[j_1],Y[{j_m}-1]}$
             \item $\cells_{p,L}[2i+1] \leftarrow \paren{i_m,j_1,i_2,j_m,X[i_m],X[i_2-1],Y[j_1],Y[j_m-1]}$
             \item $\cells_{p,R}[2i] \leftarrow \paren{i_1,j_m,i_m,j_2,X[i_1],X[i_m-1],Y[j_m],Y[j_2-1]}$
             \item $\cells_{p,R}[2i+1] \leftarrow \paren{i_m,j_m,i_2,j_2,X[i_m],X[i_2-1],Y[j_m],Y[j_2-1]}$
             \end{enumerate}
        \end{minipage}
  \item \textsl{Comment: Now $\cells_{p,R}$ and $\cells_{p,L}$ together contain the new subcells, and both arrays are sorted lexicographically.}
  \item \label{partition.merge} Merge $\cells_{p,R}$ and $\cells_{p,L}$ into an
        array $\cells^*_p$ that is sorted lexicographically.
  \item {\bf return} $\cells^*_p$.
\end{enumerate}
\end{center}
\hrule
\caption{Partitioning each cell in $\cells_p$ into four subcells.}
\label{fig:partition}
\end{figure*}
%% ...................................................................

Before we can prove that this algorithm is indeed IO-efficient, we
need to deal with one subtlety: we
need to be more specific about the exact implementation of
step~(\ref{stepb}) of the FJ-algorithm
in case the $q$th element, $x_u$, is not unique.
More precisely,  we need to specify which of the cells $C$ with $\min(C)=x_u$
are discarded and which are kept.
Similarly, we must specify which of the cells $C$ with $\max(C)=x_l$
are discarded and which are kept in step~(\ref{stepc}).
We do this as follows.

\begin{figure}[htb]\centering\small
\includegraphics[height=2in]{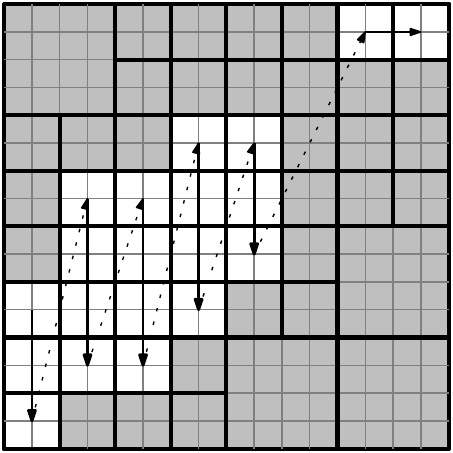}
\label{fig:active-cells}
\label{fig:staircases}
\caption{The structure of the active cells, and the order
in which they are accessed (which is the lexicographic order).
Active cells are white, discarded cells are grey.}
\end{figure}

Recall that we maintain $\cells^*_p$ in
lexicographic order. Now we can implement step~(\ref{stepb}) by removing
from $\cells^*_p$ exactly those cells whose ranks are greater than $q$ according
to this lexicographical order.  This implies that if we remove a certain cell $C$, we will
also remove all cells to the south-east of $C$ (including the ones to
the south of $C$, and the ones to the east of $C$).  We use a similar
strategy to guarantee that when we remove a cell in
step~(\ref{stepc}), we also remove all cells to its north-west.
With this implementation, the active cells have the following
properties---see also Figure~\ref{fig:active-cells}.
\begin{enumerate}
\item[(i)] All active cells with the same column index are consecutive.
\item[(ii)] The active cell with the largest row index in a given column---note that
row indices increase when going downwards in Figure~\ref{fig:active-cells}---cannot
have row index smaller than the any active cell in the column to its right. In other words,
if we consider the lowest active cells in each column
and we consider the columns from left to right, then the
the row indices of these highest active cells are non-increasing.
%Thus, active cells with
%the smallest row index in each column define a non-decreasing
%staircase from left to right.  Similarly, the active cell with the
%largest row index in a column cannot have row index smaller than that
%of the active cell with the largest row index in the column to its
%right. Thus, active cells with the largest row index in each column
%also define a non-decreasing staircase from left to right. (See
%Figure~\ref{fig:staircases}.)

\end{enumerate}

These properties are essential to get good IO-complexity of
\textsc{Partition}.

%% ...................................................................
%% ...................................................................

%% ...................................................................
\begin{lemma}
  Algorithm~\textsc{Partition} produces a lexicographically sorted
  array $\cells^*_p$ of all subcells resulting from partitioning every
  cell in $\cells_p$ into four.  \textsc{Partition} runs in $O(|\cells_p|)$ time
  and performs $O(\scan(|\cells_p|+2^p))$ IOs.
\label{le:partition}
\end{lemma}

\begin{proof}
The correctness of the algorithm directly follows from
the fact that, by definition of $\beta_1$ and $\beta_2$, the correct
values are fetched in steps~(\ref{partition.fetchx})~and~(\ref{partition.fetchy}).

To bound the running time, we
note that $\beta_1(\cdot)$ and $\beta_2(\cdot)$ can be evaluated in $O(1)$ time.
Indeed, when we evaluate e.g.~$\beta_1(j)$ for
some $j$, we know the value of $p$ such that $j=(2i-1) \cdot(n/2^p)$---this
$p$ is a parameter of~\textsc{Partition}. Given $p$, we have
$\beta_1(j) = 2^{p-1} + (j\cdot(2^p/n) +1)/2 -1$.
It follows that the running time is $O(n)$.

As for the number of IOs, all accesses
to $\cells_p$, as well as step~(\ref{partition.merge}), take
$O(\scan(|\cells_p|+2^p))$ IOs in total. Hence, it remains to argue about the accesses
to $\Xmin$, $\Xmax$, $\Ymin$, and $\Ymax$.

We first consider the accesses to $\Xmin$.
As argued earlier, the cells in $\cells_p$ have size
$(n/2^{p-1}) \times (n/2^{p-1})$, which means we need to fetch
from $\Xmin$ (a subset of) the elements $X[(2i-1) \cdot (n/2^{p-1})]$ for $1\leq i\leq 2^{p-1}$.
By the definition of $\Xmin$---see Equation~(\ref{def:permutation1})---these
elements are consecutive in $\Xmin$. Moreover, these elements are accessed
from left to right in $\Xmin$, because the
cells in $\cells_p$ are sorted in increasing order of their first
coordinate. Hence, all these accesses to $\Xmin$ take
$O(\scan(2^{p-1}))=O(\scan(2^p))$ IOs in total.
Symmetric reasoning gives the same bound on the number of accesses to $\Xmax$.

Now consider the accesses to $\Ymin$; symmetric reasoning bounds the
accesses to $\Ymax$.  Consider Figure~\ref{fig:active-cells}. The active
cells will be visited by the algorithm in lexicographic order, as
indicated in the figure. This means that the algorithm may go
back and forth in $\Ymin$. Moreover, when going back, the algorithm may \emph{jump} from
accessing some element $\Ymin[j]$ to accessing another element
$\Ymin[j']$ where $j-j' > 1$; we call $j-j'$ the \emph{length} of the jump.  Jumps are
significant because each jump may incur a cost of one IO operation.
(Jumps are also possible when accessing $\Xmin$ or $\Xmax$.
Since in $\Xmin$ and $\Xmax$ we only jump forward, this does not increase the
number of IOs there.)
Note that the elements needed within a single column of active cells, are
stored in the correct order in $\Ymin$.
(Here the term ``column'' refers to a column in the matrix of whose
cells are submatrices of size $(n/2^p)\times (n/2^p)$.)
When we step from the lowest active cell in one column
to highest active cell in the next column, however, we may jump in $\Ymin$.
% This jump can either be forwards or backwards.
Now suppose that instead of
jumping from one location to the next, we visit all intermediate
locations as well. Hence, after visiting $\Ymin[j]$, the new traversal
always proceeds to either $\Ymin[j-1]$ or $\Ymin[j+1]$. We call
such a traversal \emph{well-behaved}. Clearly the number of IOs needed
by the new traversal of $\Ymin$ is not more than the number of
traversals needed by the original traversal.

The original traversal visited $|\cells_p|$ (not necessarily distinct)
locations in $\Ymin$.  We claim that the length of the new, well-behaved traversal is
$O(|\cells^p|+2^p)$.  To show this, we must bound the total length of
all backward jumps. Consider a backward jump from the lowest active
cell in some column~$C$ to the highest active cell in the next
column~$C'$. This jump crosses a number of rows.
By properties~(i)~and~(ii) of the active cells, for each row
that is crossed, at least one of the following three
condition holds: $C$ contains an active cell in this row,
$C'$ contains an active cell in this row, or the
row will not be visited again later.  This is easily seen
to imply that the total length of all jumps is $O(|\cells^p|+2^p)$,
as claimed.

%By properties (i) and (ii) of the active cells, the total
%length of all backward jumps is no more than $|\cells_p|$.
%Indeed, a backward jump when going from one column to the next can be charged
%to the active cells in the former column.
%Furthermore, property (ii) implies that whenever we do a forward jump
%to a location $\Ymin[j]$, subsequent forward jumps must start at some
%location $\Ymin[j']$ with $j \le j'$. Hence, the total length of all
%forward jumps is at most the total size of the part of $\Ymin$ that we
%are visiting in the $p$-th iteration, which is $2^p$.  This proves our
%claim that the length of the new traversal is
%$O(|\cells^p|+2^p)$.

It remains to observe that, assuming $M\ge 2B$---that is, assuming at
least two blocks fit in the cache---any well-behaved traversal of
length $L$ needs $\scan(L)$ IOs. Indeed, suppose we need to read a new
block when we step from $\Ymin[i]$ to $\Ymin[i+1]$. Then we read the block starting
at $\Ymin[i+1]$ and can keep the block ending at $\Ymin[i]$ in cache.
Hence, at least $B-1$ more forward steps or at least
$B$ backward steps are needed before another block needs to be read.
We conclude that the number of IOs performed in accessing $\Ymin$ (and,
similarly, $\Ymax$) is $O(\scan(|\cells^p|+2^p)$, which finishes the proof
for the number of IOs.
\end{proof}
%% ...................................................................

%% ..........................................................
\begin{theorem}
  There exists a cache-oblivious implementation of the matrix
  selection algorithm of Frederickson and Johnson for sorted $X+Y$
  matrices using $O(\scan(n))$ IOs and $O(n)$ time, where $n$
  is the maximum of the lengths of $X$ and $Y$.
\end{theorem}

\begin{proof}
By Lemma~\ref{lemma:permutecost}, the computation of the arrays
$\Xmin$, $\Xmax$, $\Ymin$, and $\Ymax$ takes $O(\scan(n))$ IOs and $O(n)$ time.
Now consider the main algorithm.
Frederickson and Johnson~\cite{frederickson84generalized} proved that
$|\cells^p|$, the number of active cells in the beginning of the $p$th
iteration, is $O(2^p)$.  By Lemmas~\ref{le:standard-selection},
\ref{le:reorder}, and~\ref{le:partition}, this implies that the total
IO-cost is bounded by
\[
  \sum_{p=1}^{\lg n} O(\scan(2^p)) = O(\scan(n)).
\]
Since the subroutine \textsc{Partition} runs in $O(|\cells_p|)$,
the running time
of the main algorithm is unchanged from the original FJ-algorithm, which runs
in~$O(n)$ time.
\end{proof}

%%%%%%%%%%%%%%%%%%%%%%%%%%%%%%%%%%%%%%%%%%%%%%%%%%%%%%%%%%%%%%%%%%%%%%%%%%%%%
%%%%%%%%%%%%%%%%%%%%%%%%%%%%%%%%%%%%%%%%%%%%%%%%%%%%%%%%%%%%%%%%%%%%%%%%%%%%%
%%%%%%%%%%%%%%%%%%%%%%%%%%%%%%%%%%%%%%%%%%%%%%%%%%%%%%%%%%%%%%%%%%%%%%%%%%%%%
%%%%%%%%%%%%%%%%%%%%%%%%%%%%%%%%%%%%%%%%%%%%%%%%%%%%%%%%%%%%%%%%%%%%%%%%%%%%%
\section{Conclusion}

In this paper, we gave an IO-efficient cache-oblivious version of the
classical matrix selection algorithm of Frederickson and Johnson for
selecting a rank-$k$ element in an $n \times n$ matrix given
succinctly in the form $A := X+Y$.

If the matrix $A$ is not square---that is, if its dimensions were $m
\times n$ where $m < n$---then a different approach seems to be
required to make the matrix selection algorithm IO-efficient.
One would like to obtain an IO-cost of
\[
  O\paren{\frac{m}{B} \log_B \frac{2n}{m}}.
\]
However, we already spend $O((m+n)/B)$ IOs
in permuting the input arrays as a pre-processing step, which
dominates the IO-cost of the subsequent algorithm. It seems
difficult to avoid the high IO-cost of permuting both input arrays so
that they can be accessed IO-efficiently.  A completely new algorithm
may be necessary to achieve IO-optimal matrix selection in sorted
$X+Y$ matrices that are not square.

%% ##################################################################
\bibliographystyle{plain}
% \bibliography{select,io-efficient}

\newcommand{\etalchar}[1]{$^{#1}$}

\end{document}